\documentclass{elsart}
\usepackage{graphicx,amssymb}
\usepackage{epsfig}
\usepackage{setspace}

\begin{document}
\begin{frontmatter}

\title{A recursive approach for the finite element computation of waveguides}

\author[Lami]{Denis Duhamel}

\address[Lami]{Universit\'e Paris-Est, UR Navier, \\
Ecole Nationale des Ponts et Chauss\'ees, \\
6 et 8 Avenue Blaise Pascal, \\
Cit\'{e} Descartes, Champs sur Marne,\\
77455 Marne la Vall\'{e}e, cedex 2, France\\
Tel: 33 1 64 15 37 28 \\
Fax: 33 1 64 15 37 41 \\
email{\nobreakspace}: duhamel@lami.enpc.fr}

\vspace{5cm}

Number of pages : 23

Number of figures : 11

Number of tables : 1

\date{}
\newpage

\begin{abstract}
The finite element computation of structures such as waveguides can
lead to heavy computations when the length of the structure is large
compared to the wavelength. Such waveguides can in fact be seen as
one-dimensional periodic structures. In this paper a simple
recursive method is presented to compute the global dynamic
stiffness matrix of finite periodic structures. This allows to get
frequency response functions with a small amount of computations.
Examples are presented to show that the computing time is of order
$\log_2 N$ where $N$ is the number of periods of the waveguide.
\end{abstract}

\end{frontmatter}

\newpage

\section{Introduction}

\begin{figure}
  \begin{center}
    \epsfig{file=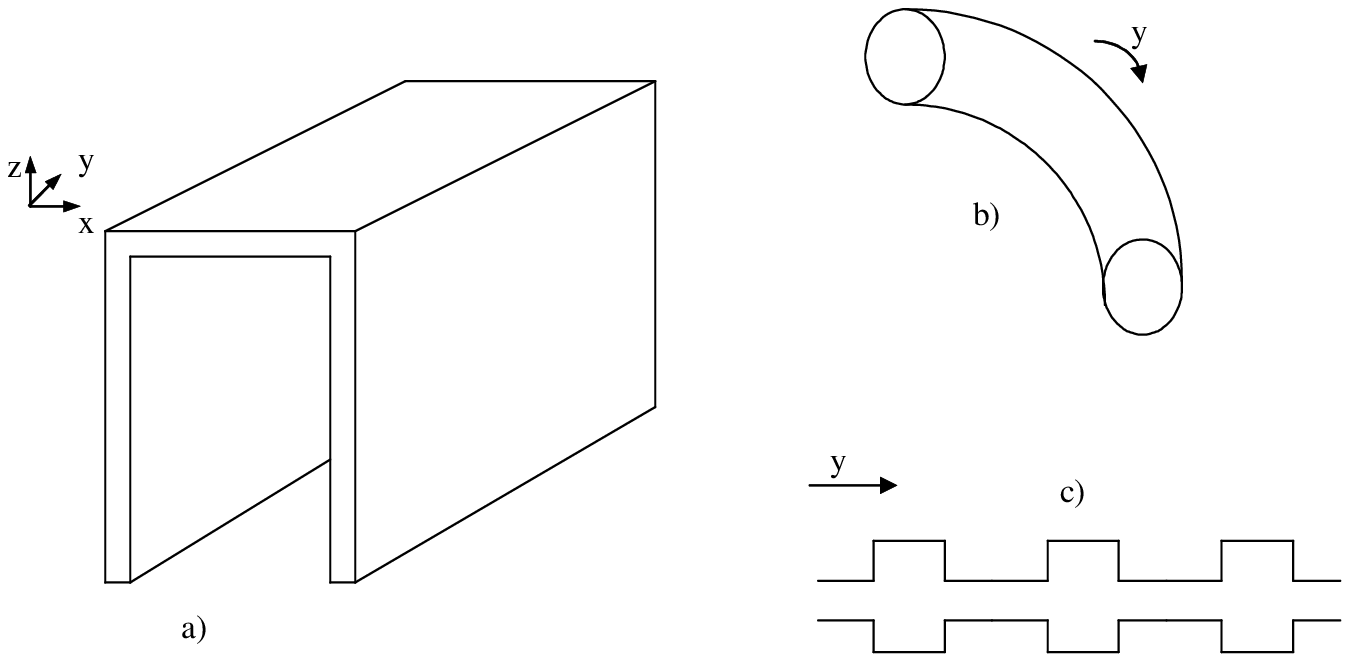,scale=.7}
    \caption{Examples of waveguide structures.\label{fig1}}
  \end{center}
\end{figure}

We study here the computation of structures considered as
waveguides, as shown in figure \ref{fig1}, with symmetries which can
be a translation (a), a rotation (b) or a periodicity (c). 
Thus these waveguides can be uniform or periodic.
The vibration of such waveguides has been the topic of much research.
One can find analytical or finite element models of waveguides
and people are generally interested by the computation of wave propagations 
and dispersion curves or by the determination of the frequency response
functions.
A first approach considers structures with constant cross sections
as the cases a) and b) of figure 1. For example, \cite{Von1} and
\cite{Bea1} used a wave approach to study the vibrations of
structural networks composed of simple uniform beams, and solved for
the dynamics of individual elements and of the junctions between
elements by analytical methods. The efficiency is greatly improved
compared to FE methods as a beam can be modelled using only a single
element.

The first numerical approaches were proposed by \cite{Don1,Aal1}
to approximate the cross-sectional deformations by finite elements.
The authors of \cite{Gav1,Gry1} applied similar ideas to the calculation
of wave propagations in rails using a finite element model of the
cross-section of a rail. 
They then calculated dispersion relations and accelerances. 
Dispersion relations for elastic waves in helical waveguides 
were also considered by \cite{Tre1}.
For general waveguides with a complex cross-section,
the displacements in the cross-section can be described by the
finite element method while the variation along the axis of symmetry
is expressed as a wave function. Following these ideas,
\cite{Fin1,Fin2,Nil1,Bir1,Bir2,Fin3} developed the spectral finite
element approach. This leads to efficient computations of dispersion
relations and transfer functions but special elements need to be
developed for each element type. This makes the connection with the
standard use of the finite element method difficult and does not
allow the benefits of powerful existing finite element software to
be exploited.
Similar techniques were also developed by \cite{Bar1,Mar1} 
for the computation of dispersion relations in damped waveguides.

More general waveguides can be studied by considering periodic
structures. Numerous works provided interesting theoretical insights
in the behaviour of these structures, see for instance the work of
\cite{Bri1} and the review paper by \cite{Mead1}. Mead also
presented a general theory for wave propagation in periodic systems
in \cite{Mead2,Mead3,Mead4}. He showed that the solution can be
decomposed into an equal number of positive and negative-going
waves. The approach is mainly based on Floquet's principle or the
transfer matrix and the objective it to compute propagation
constants relating the forces and displacements on the two sides of a
cell (a single period) and the waves associated to these constants.
For complex structures FE models are used for the computation of the
propagation constants and waves. The final objective is to compute
dispersion relations to use them in energetic methods, see
\cite{Hou1,Mac1,Mac2,Men1,Men2}. In \cite{Duh1} the general dynamic
stiffness matrix for a periodic structure was found from the
propagation constants and waves. It leads to a matrix linking the
extreme sides of the structure and allows to compute transfer
functions in the structure.

The last approach is purely computational and uses rotational and
cyclic symmetries to solve the problem by a decomposition of the
displacements in cosine and sine functions. It is thus possible to
find the transfer functions or modal shapes for periodic structures.
A review of the current practises can be found in \cite{Wan1}. These
methods allow computing the frequency response functions in a number
of operations proportional to the number of cells in the structure.
This paper develops this last approach and presents a recursive
method to calculate the forced response of structures such as those
illustrated in figure 1. A section of the waveguide is modelled
using conventional FE methods, using a commercial FE package. The
resulting mass, stiffness and damping matrices are then
post-processed to give the dynamic stiffness matrix of the cell.
Then a recursive method is applied to compute the global dynamic
stiffness matrix of the whole waveguide and finally the transfer
functions in the structure. 

This paper presents a different approach from the previously 
published paper \cite{Duh1}.
Both papers aim at computing the global dynamic stiffness matrix of a N cells structure.
But in \cite{Duh1} waves in a period were computed and from these waves
the dynamic stiffness matrix of a complete structure was obtained
with a computational cost independent of the number of periods in the structure.
However the computation of the waves can be time consuming when the number of dofs 
in a section is large because this needs the computation of eigenvalues
of non symmetric matrices.
On the contrary, in the present approach no wave needs to be computed 
and the global dynamic stiffness matrix is obtained by products and inverses 
of matrices with the same dimensions as the dynamic stiffness matrix of a cell.
Then, the frequency response functions can be obtained easily without the computation of any wave.

The paper is divided into two parts. 
In the first part the recursive method for the finite element analysis
of periodic structures is presented. In the second part two examples
consisting in a beam and a plate are described before the
conclusion.

%%%%%%%%%%%%%%%%%%%%%%%%%%%%%%%%%%%%%%%%%%%%%%%%%%%%%%%%%%%%%%%%%%%%%%%%%%%%%%%%

\section{Finite element analysis of periodic structures}

\begin{figure}
  \begin{center}
    \epsfig{file=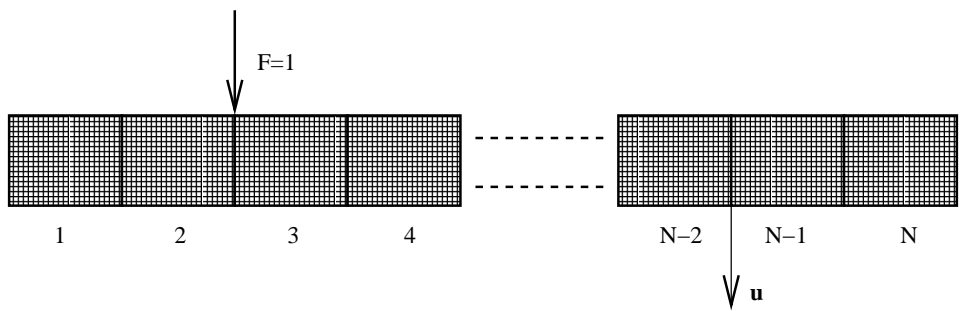,scale=.5}
    \caption{Periodic structure made of N cells with an excitation
    by a force $\mathbf{F}$.\label{fig2}}
  \end{center}
\end{figure}

Consider a periodic structure, as shown in figure \ref{fig2}, which
is made of a large number $N$ of cells. We are interested by the
computation of the frequency response function for a point force
excitation $F=1$ somewhere in the structure and a response
$u$ at another point. We propose here an efficient method
to compute this function by using a recursive approach to get the
dynamic stiffness matrix of different sets of cells.

\subsection{Behaviour of a cell}

\begin{figure}
  \begin{center}
    \epsfig{file=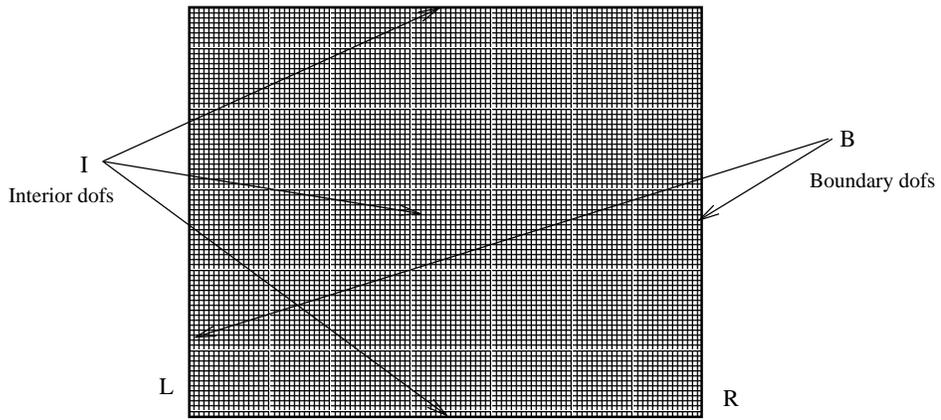,scale=.5}
    \caption{Interior and boundary dofs for a single cell.\label{fig3}}
  \end{center}
\end{figure}

Consider first the case of only one cell. The discrete dynamic
equation of a cell obtained from a FE model at a frequency $\omega$
and for the time dependence $e^{-i\omega t}$ is given by:
\begin{equation}
  (\mathbf{K} - i\omega\mathbf{C} - \omega^2\mathbf{M})\mathbf{q} =
  \mathbf{f}
\label{eq01}
\end{equation}

where $\mathbf{K}$, $\mathbf{M}$ and $\mathbf{C}$ are the stiffness,
mass and damping matrices, respectively, $\mathbf{f}$ is the loading
vector and $\mathbf{q}$ the vector of the degrees of freedom (dofs).
A viscous damping is considered here but the same results could be
obtained with other damping models.
Introducing the dynamic stiffness
matrix $\widetilde{\mathbf{D}} = \mathbf{K} - i\omega\mathbf{C} -
\omega^2\mathbf{M}$, decomposing the dofs into boundary $(B)$ and
interior $(I)$ dofs as shown in figure \ref{fig3}, and assuming that
there are no external forces on the interior nodes, result in the
following equation:
\begin{equation}
  \left [
    \begin{array}{cc}
      \widetilde{\mathbf{D}}_{BB} & \widetilde{\mathbf{D}}_{BI}\\
      \widetilde{\mathbf{D}}_{IB} & \widetilde{\mathbf{D}}_{II}
    \end{array}
  \right ]
  \left [
    \begin{array}{c}
      \mathbf{q}_{B}\\
      \mathbf{q}_{I}
    \end{array}
  \right ]
  =
  \left [
    \begin{array}{c}
      \mathbf{f}_{B}\\
      \mathbf{0}
    \end{array}
  \right ]
  \label{eq02}
\end{equation}

The interior dofs can be eliminated using the second row of
equation (\ref{eq02}), which results in
\begin{equation}
  \mathbf{q}_I =
  -\widetilde{\mathbf{D}}_{II}^{-1}\widetilde{\mathbf{D}}_{IB}\mathbf{q}_B
\label{eq03}
\end{equation}

The first row of equation (\ref{eq02}) becomes
\begin{equation}
  \mathbf{f}_B =
\left(\widetilde{\mathbf{D}}_{BB}-\widetilde{\mathbf{D}}_{BI}\widetilde{\mathbf{D}}_{II}^{-1}\widetilde{\mathbf{D}}_{IB}\right
)\mathbf{q}_B \label{eq04}
\end{equation}

It should be noted that only boundary dofs are considered in the
following. The cell is assumed to be meshed with an equal number of
nodes on their opposite sides. The boundary dofs for one cell are
decomposed into left $(L)$ and right $(R)$ dofs as shown in figure
\ref{fig3}. Thus, equation (\ref{eq04}) is rewritten as
\begin{equation}
  \left [
    \begin{array}{c}
      \mathbf{f}_L\\
      \mathbf{f}_R
    \end{array}
  \right ] =
  \left [
    \begin{array}{cc}
      \mathbf{D}_{LL}^{(1)} & \mathbf{D}_{LR}^{(1)} \\
      \mathbf{D}_{RL}^{(1)} & \mathbf{D}_{RR}^{(1)} \\
    \end{array}
  \right ]
  \left [
    \begin{array}{c}
      \mathbf{q}_L\\
      \mathbf{q}_R
    \end{array}
  \right ]
= \mathbf{D}^{(1)}
  \left [
    \begin{array}{c}
      \mathbf{q}_L\\
      \mathbf{q}_R
    \end{array}
  \right ]
  \label{eq06}
\end{equation}
where $\mathbf{D}^{(1)}$ is the dynamic stiffness matrix of a single cell.
This matrix is symmetric if the matrices $\mathbf{K}$, $\mathbf{M}$ and $\mathbf{C}$
in relation (\ref{eq01}) are symmetric.

\subsection{Computation of reduced dynamic stiffness matrices}

\begin{figure}
  \begin{center}
    \epsfig{file=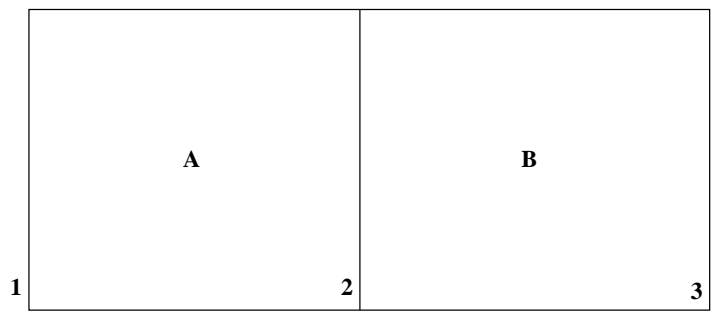,scale=.5}
    \caption{Structure with two cells and three sections.\label{fig4}}
  \end{center}
\end{figure}

Consider a structure made of two cells with the respective dynamic
stiffness matrices denoted by $\mathbf{A}$ and $\mathbf{B}$ as in
figure \ref{fig4}. We propose to remove the internal degrees of
freedom at the boundary between the two cells to compute the dynamic
stiffness matrix, denoted $\mathbf{D}^{(2)}$, relating the degrees of
freedom (dofs) in the first section of $A$ and the last section of
$B$. The dynamic stiffness matrix of the substructure with two cells
is computed by
\begin{equation}
  \left [
    \begin{array}{c}
      \mathbf{f}_1\\
      \mathbf{f}_2\\
      \mathbf{f}_3
    \end{array}
  \right ]
= \left [
    \begin{array}{ccc}
      \mathbf{A}_{LL} & \mathbf{A}_{LR} & 0 \\
      \mathbf{A}_{RL} & \mathbf{A}_{RR}+\mathbf{B}_{LL} & \mathbf{B}_{LR} \\
      0 & \mathbf{B}_{RL} & \mathbf{B}_{RR} \\
    \end{array}
  \right ]
  \left [
    \begin{array}{c}
      \mathbf{q}_1\\
      \mathbf{q}_2\\
      \mathbf{q}_3
    \end{array}
  \right ]
\label{eq07}
\end{equation}
As there is no load on the interior section, one gets $\mathbf{f}_2=0$ and
\begin{equation}
\mathbf{q}_2
= -\left(\mathbf{A}_{RR}+\mathbf{B}_{LL}\right)^{-1}
\left(\mathbf{A}_{RL}\mathbf{q}_1+\mathbf{B}_{LR}\mathbf{q}_3\right)
\end{equation}
The global dynamic stiffness matrix of the two cells structure is thus
\begin{eqnarray}
  \left [
    \begin{array}{c}
      \mathbf{f}_1\\
      \mathbf{f}_3
    \end{array}
  \right ]
& = & \left [
    \begin{array}{cc}
  \mathbf{A}_{LL} - \mathbf{A}_{LR}\left(\mathbf{A}_{RR}+\mathbf{B}_{LL}\right)^{-1}\mathbf{A}_{RL} &
  - \mathbf{A}_{LR}\left(\mathbf{A}_{RR}+\mathbf{B}_{LL}\right)^{-1}\mathbf{B}_{LR} \\
  - \mathbf{B}_{RL}\left(\mathbf{A}_{RR}+\mathbf{B}_{LL}\right)^{-1}\mathbf{A}_{RL} &
  \mathbf{B}_{RR} - \mathbf{B}_{RL}\left(\mathbf{A}_{RR}+\mathbf{B}_{LL}\right)^{-1}\mathbf{B}_{LR} \\
    \end{array}
  \right ]
  \left [
    \begin{array}{c}
      \mathbf{q}_1\\
      \mathbf{q}_3
    \end{array}
  \right ] \nonumber \\
& = & \left [
    \begin{array}{ccc}
      \mathbf{D}_{LL}^{(2)} & \mathbf{D}_{LR}^{(2)} \\
      \mathbf{D}_{RL}^{(2)} & \mathbf{D}_{RR}^{(2)}
    \end{array}
  \right ]
  \left [
    \begin{array}{c}
      \mathbf{q}_1\\
      \mathbf{q}_3
    \end{array}
  \right ] \nonumber \\
& = & \mathbf{D}^{(2)}
  \left [
    \begin{array}{c}
      \mathbf{q}_1\\
      \mathbf{q}_3
    \end{array}
  \right ]
\label{eq08}
\end{eqnarray}
This defines the matrix $\mathbf{D}^{(2)}$ which relates the forces and
displacements dofs at the extreme sections of the two-cells
structure. 
It can be easily checked that if the matrices $\mathbf{A}$ and $\mathbf{B}$
are symmetric, the resulting matrix $\mathbf{D}^{(2)}$ of relation (\ref{eq08})
is also symmetric.
The operation of removing the interior dofs is now
denoted by $\left\{.,.\right\}$ such that we can write

\begin{equation}
\mathbf{D}^{(2)} = \left\{\mathbf{A},\mathbf{B}\right\}
\end{equation}

\subsection{Case of general structures}

\begin{figure}
  \begin{center}
    \epsfig{file=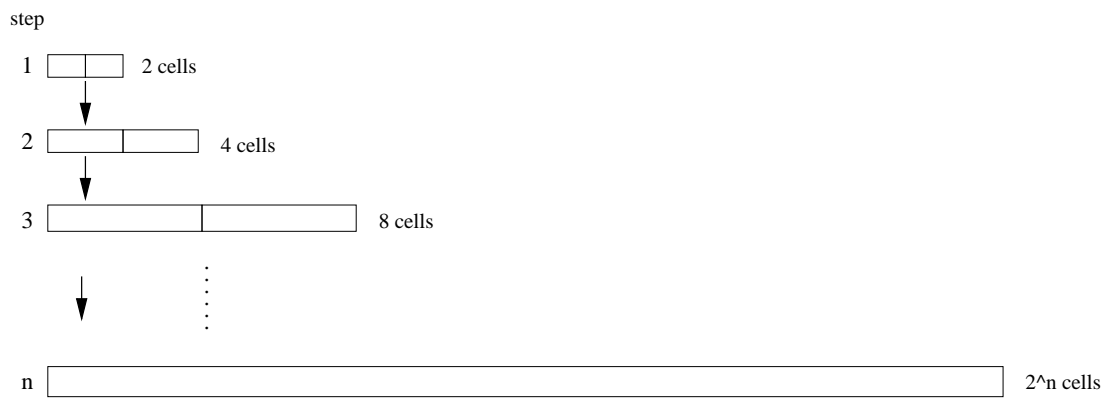,scale=.5}
    \caption{Structure at each step.\label{fig5}}
  \end{center}
\end{figure}

Consider now a structure without internal load and made of $2^n$
cells. We propose to recursively remove the internal degrees of
freedom between adjacent cells to compute the dynamic stiffness
matrix, denoted $\mathbf{D}^{(2^n)}$, relating the degrees of freedom
in sections $1$ and $2^n$. Consider firstly a structure with two identical
cells. From the precedent analysis, one sees that its dynamic
stiffness matrix is given by
$\mathbf{D}^{(2)}=\left\{\mathbf{D}^{(1)},\mathbf{D}^{(1)}\right\}$. Repeating
the process (see an illustration in figure \ref{fig5}), one gets the
dynamic stiffness matrix of the structure with $2^n$ cells in $n$
steps by the recursive relation

\begin{equation}
\mathbf{D}^{(2^n)} = \left\{\mathbf{D}^{(2^{n-1})},\mathbf{D}^{(2^{n-1})}\right\}
\end{equation}

This matrix is such that
\begin{equation}
  \left [
    \begin{array}{c}
      \mathbf{f}_1\\
      \mathbf{f}_{2^n}
    \end{array}
  \right ]
= \left [
    \begin{array}{ccc}
      \mathbf{D}_{LL}^{(2^n)} & \mathbf{D}_{LR}^{(2^n)} \\
      \mathbf{D}_{RL}^{(2^n)} & \mathbf{D}_{RR}^{(2^n)}
    \end{array}
  \right ]
  \left [
    \begin{array}{c}
      \mathbf{q}_1\\
      \mathbf{q}_{2^n}
    \end{array}
  \right ]
= \mathbf{D}^{(2^n)}
  \left [
    \begin{array}{c}
      \mathbf{q}_1\\
      \mathbf{q}_{2^n}
    \end{array}
  \right ]
\label{eq09}
\end{equation}

In cases where the structure is not composed of a number of cells
which equals a power of two, one can modify the previous procedure
using the binary representation of the total number of cells N.
Consider the example where $N =11 = 1011_b$ in binary
representation. One calculates first the dynamic stiffness matrix
for a structure with 8 cells, then this structure is assembled with
a structure made of 2 cells which has been computed during the
computation of the 8 cells structure. Finally the resulting matrix
is assembled with a one cell matrix. The approach can be resumed by
\begin{equation}
\mathbf{D}^{(11)} =
\left\{\left\{\mathbf{D}^{(8)},\mathbf{D}^{(2)}\right\},\mathbf{D}^{(1)}\right\}
\end{equation}
in which the matrices $\mathbf{D}^{(2^n)}$ are computed by the method
presented before. The elimination of the interior degrees of
freedom gives the matrix linking the forces and displacements
degrees of freedom in sections $1$ and $N$. One can notice that the
final matrix is computed with a number of operations of order $\log_2
N$, thus saving a huge number of computations when we compare to a
standard approach in which all the matrices of the cells are
assembled into a global matrix.

Using this method it is easy to compute the matrices for the parts
of the structure respectively on the left and on the right of the
force. Assembling these two matrices, applying the appropriate
boundary conditions on the first and last sections, one gets the
final linear system. This system has a number of dofs which equals 
approximately three times the number of dofs in a section.

\section{Examples}

\subsection{Beam structure}

\begin{figure}
  \begin{center}
    \epsfig{file=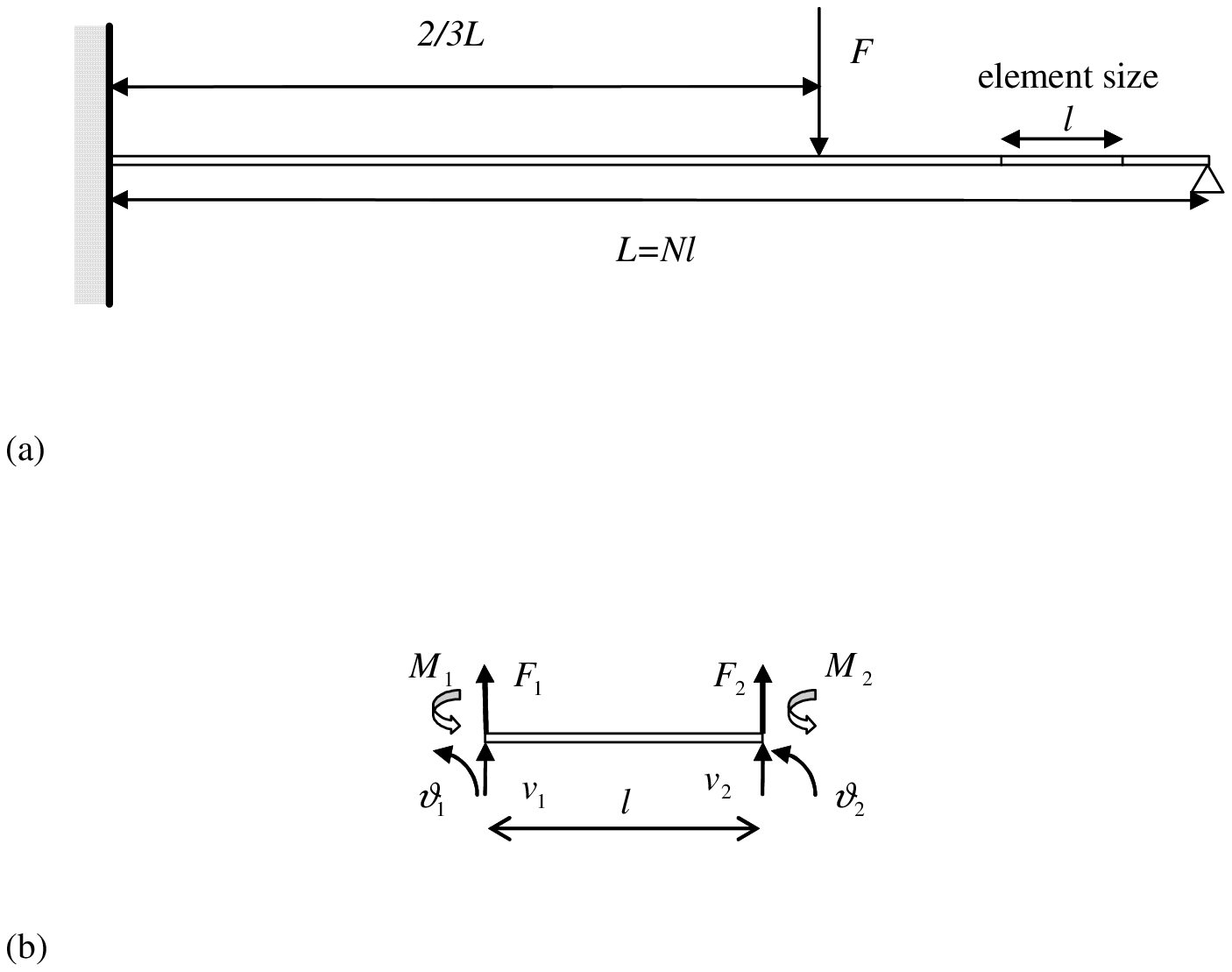,scale=.7}
    \caption{Beam structure (a) and beam element (b).\label{fig6}}
  \end{center}
\end{figure}

\begin{figure}
  \begin{center}
    \epsfig{file=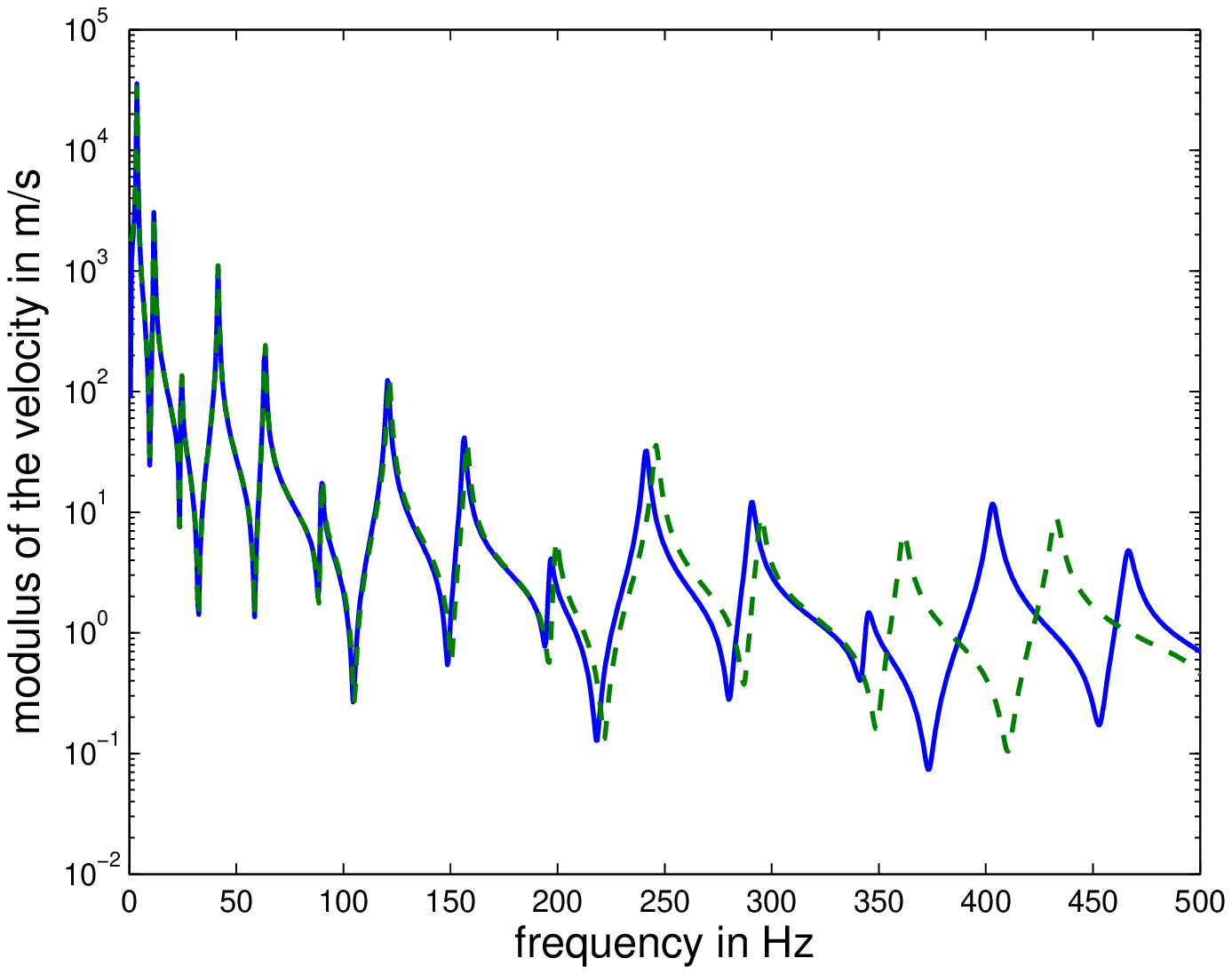,scale=.7}
    \epsfig{file=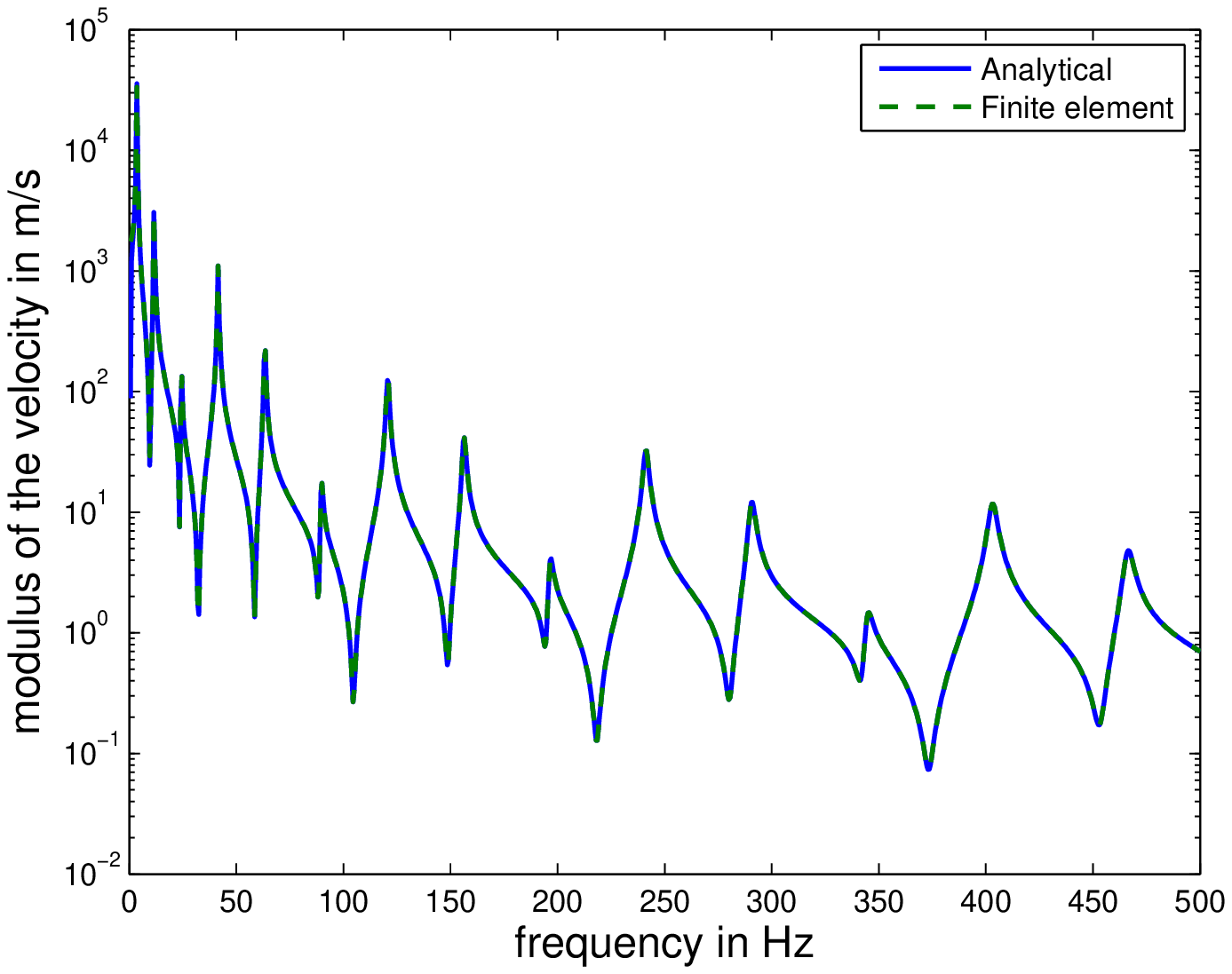,scale=.7}
    \caption{Frequency response functions for a beam with $2\times8$ elements (upper graph)
and $2\times1024$ elements (lower graph): $\line(1,0){20}$
analytical solution, $---$ finite element solution. 
The position of the excitation point is shown in figure \ref{fig6}
and the response is computed at the same point. \label{fig7}}
  \end{center}
\end{figure}

In the first example, we consider the beam shown in figure
\ref{fig6} which is made of elements with four dofs. The stiffness
and mass matrices of an element of length $l$ are given by
\begin{equation}
\mathbf{K}_e = \frac{EI}{l^3}
  \left [
    \begin{array}{cccc}
    12 & 6l & -12 & 6l \\
    6l & 4l^2 & -6l & 2l^2 \\
    -12 & -6l & 12 & -6l \\
    6l & 2l^2 & -6l & 4l^2
    \end{array}
  \right ]
\end{equation}

\begin{equation}
\mathbf{M}_e = \frac{\rho Sl}{420}
  \left [
    \begin{array}{cccc}
    156 & 22l & 54 & -13l \\
    22l & 4l^2 & 13l & -3l^2 \\
    54 & 13l & 156 & -22l \\
    -13l & -3l^2 & -22l & 4l^2
    \end{array}
  \right ] \nonumber
\end{equation}
Here, $E$ is the Young's modulus, $I$ the second moment of area,
$\rho$ the density of the material and $S$ the cross-sectional area of the beam.
Using the previous approach, one can compute the dynamic stiffness
matrices for the sections on the left and on the right of the force.
The damping matrix is obtained by using a complex Young modulus such
that $E=E_0(1+i\eta)$ with $\eta = 0.01$ leading to a hysteretic
damping instead of the viscous one such that $\tilde{\mathbf{D}} =
(1+i\eta)\mathbf{K}-\omega^2\mathbf{M}$.
The beam is made of steal such that $E_0=2\times 10^{11}\mathrm{Pa}$,  
$\rho=7800 \mathrm{kg/m^3}$, the length of the beam $L=1\mathrm{m}$, 
$I=8.33\times 10^{-14}\mathrm{m}^4$ and $S=10^{-6}\mathrm{m}^2$.
The matrix for the whole
beam is obtained by assembling the matrices of the left and right
parts of the beam on each sides of the force. Taking into account
the fixed displacement boundary conditions by removing the
corresponding degrees of freedom in the global matrix results in the
final system. Figure \ref{fig7} presents the frequency response
functions for structures with respectively 8 and 1024 elements in
each part of the beam. It can be seen that 8 elements are not
sufficient to compute accurately the solution while 1024 elements
lead to a very good result.

\begin{figure}
  \begin{center}
    \epsfig{file=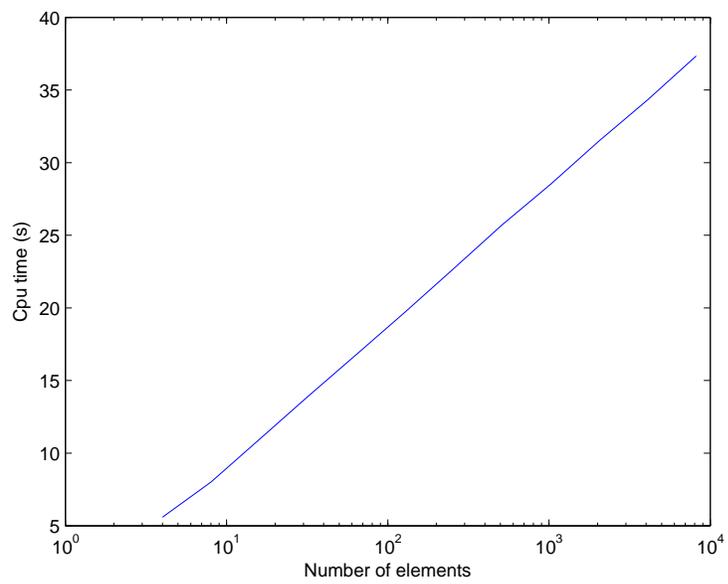,scale=.7}
    \caption{Computing time versus the number of cells in the beam.\label{fig8}}
  \end{center}
\end{figure}

In figure \ref{fig8}, the cpu time of the computation is plotted
versus the global number of elements in the beam. The computation of
10000 points in frequency is made for each mesh of the beam and the
largest mesh has $2\times 4196$ elements for the complete beam. A
linear behavior of the cpu time versus the logarithm of the
number of elements can be seen as expected.

\subsection{Plate}

\begin{figure}
  \begin{center}
    \epsfig{file=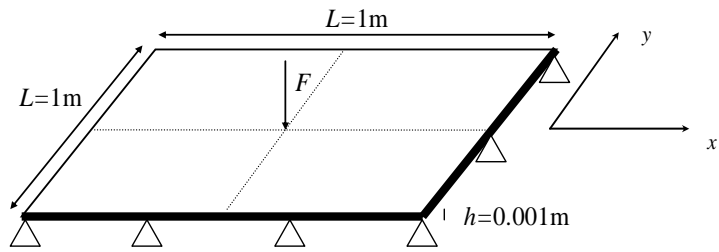,scale=.7}
    \caption{Plate example.\label{fig9}}
  \end{center}
\end{figure}

\begin{figure}
  \begin{center}
    \epsfig{file=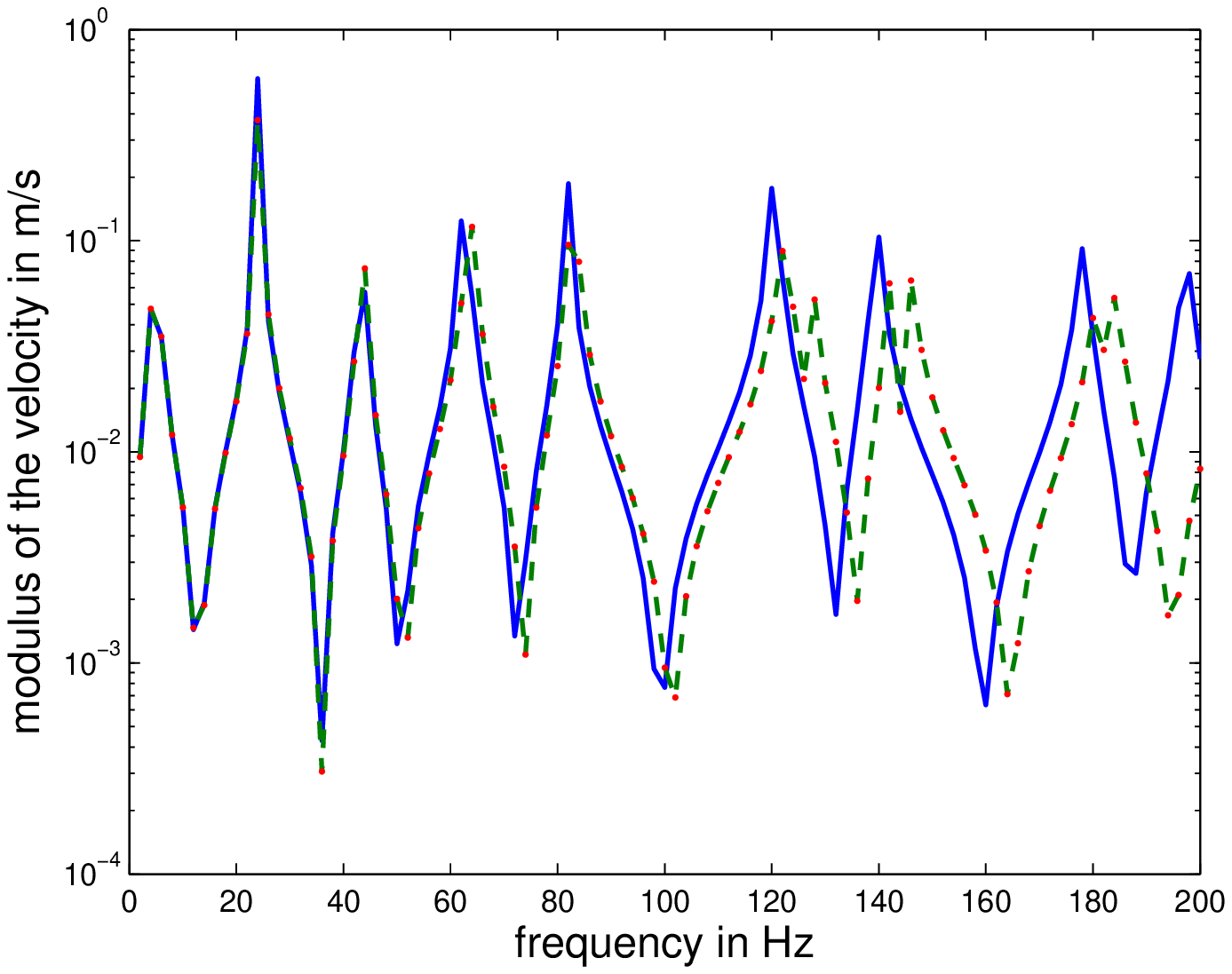,scale=.7}
    \epsfig{file=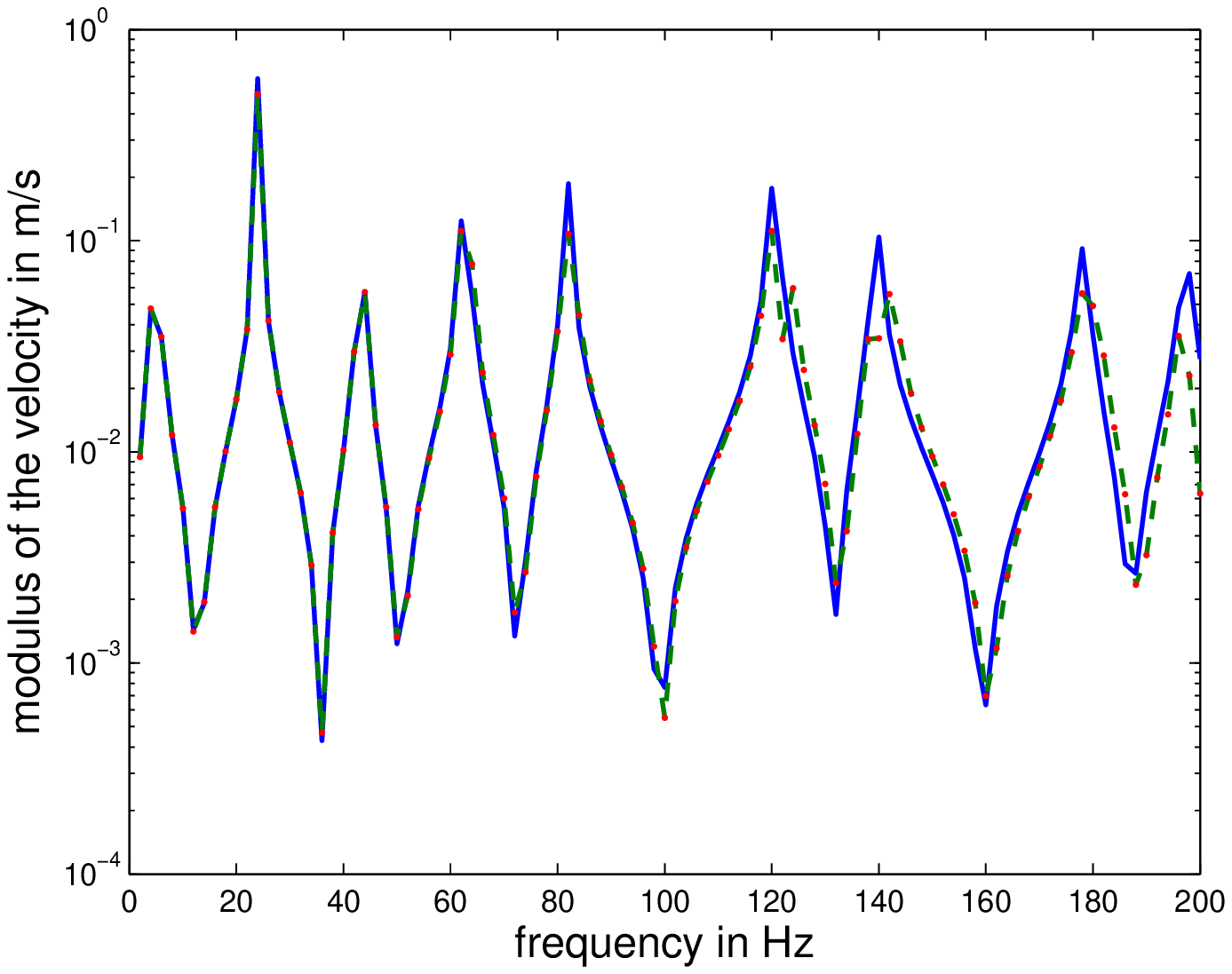,scale=.7}
    \caption{Frequency response functions for a plate with $2\times 16$ cells (upper graph)
and $2\times 256$ cells (lower graph): $\line(1,0){20}$ analytical
solution, $---$ recursive finite element solution and $.\ .$ standard finite element solution. 
The excitation is located at the center of the plate 
as shown in figure \ref{fig9} and the response is computed at the same point. \label{fig10}}
  \end{center}
\end{figure}

Consider now the plate shown in figure \ref{fig9}. The mesh of a cell
is obtained by Abaqus and consists in 50 elements of size $Ly/50
\times Lx/2^n$ where $n$ is the number of cells along the direction
$x$. The mass and stiffness matrices are produced by Abaqus then
there are loaded in Matlab and the precedent procedure allows 
computing the displacement for a load at the centre of the plate.
Information on the size are given in figure \ref{fig9}
and the plate is still made of steal. The boundary
conditions are simply supported on all sides. Figure \ref{fig10}
presents the frequency response functions for structures with
globally 32 and 512 cells. It can be seen, as for the beam, that 32
cells are not sufficient to compute accurately the solution while
512 cells lead to much better results. For high frequencies a
discrepancy with the analytical solution can still be seen. It has
been checked that the result can be considerably improved 
by taking 100 elements instead of 50 along the direction $y$.
The figures also present the results from the standard finite element approach
obtained by assembling the elementary matrices for each period
and solving the global linear system.
It can be seen that both finite elements computations yield identical results.

\begin{figure}
  \begin{center}
    \epsfig{file=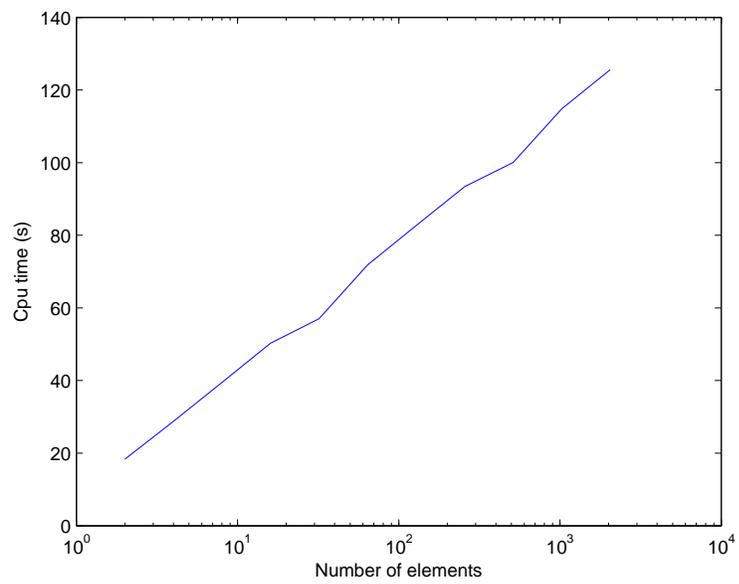,scale=.7}
    \caption{Computing time versus the number of cells in the plate.\label{fig11}}
  \end{center}
\end{figure}

In figure \ref{fig11}, the cpu time of the computation is plotted
versus the global number of cells in the plate. The computation of
100 points in frequency is made for each mesh of the plate and the
largest mesh has 2048 cells along $x$ for the complete plate. 
Once again a linear behavior of the cpu time versus 
the logarithm of the number of cells can be observed.
The computing times for the recursive and standard finite element methods 
are compared in table \ref{tab01}.
It can be seen that the recursive method is a little slower than the standard method 
for a number of periods lower than 16.
For a larger number of periods the recursive method tends to be more and more efficient 
as the number of periods increases.

\section{Conclusion}

A method has been described to compute frequency response
functions for waveguide structures with periodic or homogeneous
sections. The proposed method allows computing the solution in a
time proportional to the logarithm of the numbers of cells in the
structure. This simple recursive method can be applied for any type
of one-dimensional periodic waveguides. It could be used in the
future for the computation of complex structures such as tyres.

\newpage
\clearpage

\newpage
\clearpage

\newpage
\clearpage

\newpage
\clearpage

\newpage
\clearpage

\newpage
\clearpage

\newpage
\clearpage

\newpage
\clearpage

\newpage
\clearpage

\newpage
\clearpage

\newpage
\clearpage

\newpage
\begin{center}
\begin{table}
\caption{Computing times for recursive and standard FEM methods.}
   \begin{tabular}{|c|c|c|} \hline
number of periods & recursive method (s) & classical fem (s) \\ \hline
2 & 18.3 & 14.4 \\\hline
4 & 28.8 & 20.1 \\ \hline
8 & 39.5 & 33.7 \\ \hline
16 & 50.3 & 69.2 \\ \hline
32 & 57.0 & 164.3 \\ \hline
64 & 71.8 & 435.0 \\ \hline
128 & 82.7 & 1295.0 \\ \hline
256 & 93.4 & 4093.1 \\ \hline
512 & 100.0 & 13901.4 \\ \hline
1024 & 114.8 & 49271.1 \\ \hline
2048 & 125.6 & 184397.3 \\ \hline 
\end{tabular}
\label{tab01}
\end{table}
\end{center}

\end{document}